\documentclass[11pt,twoside]{article}

\usepackage[a4paper, left=25mm, right=25mm, top=20mm, bottom=40mm]{geometry}
\usepackage{bm,amssymb,amsfonts,amsmath}  % bold math symbols, AMStex
\usepackage{graphicx}
\usepackage{tikz,lipsum}
\usetikzlibrary{shapes}
\usetikzlibrary{positioning}
\usetikzlibrary{calc}
\usetikzlibrary{math}
\usepackage{tcolorbox}  
\tcbuselibrary{skins,breakable}  % allows tcolorboxes to break across pages
\newcommand\AIC{\textsc{aic}}
\newcommand\BIC{\textsc{bic}}
\newcommand{\matA}{\mathbb{A}}
\newcommand{\matB}{\mathbb{B}}
\newcommand{\matD}{\mathbb{D}}
\newcommand{\matE}{\mathbb{E}}
\newcommand{\matF}{\mathbb{F}}
\newcommand{\matH}{\mathbb{H}}
\newcommand{\matI}{\mathbb{I}}
\newcommand{\matL}{\mathbb{L}}
\newcommand{\matP}{\mathbb{P}}
\newcommand{\matS}{\mathbb{S}}
\newcommand{\matX}{\mathbb{X}}

\newcommand{\smK}{{\scriptscriptstyle K}}
\newcommand{\smL}{{\scriptscriptstyle L}}
\newcommand{\smM}{{\scriptscriptstyle M}}
\newcommand{\smN}{{\scriptscriptstyle N}}
\newcommand{\smS}{{\scriptscriptstyle S}}

\newcommand{\bmf}{{\bm{f}}}
\newcommand{\bmv}{{\bm{v}}}
\newcommand{\bmx}{{\bm{x}}}
\newcommand{\bmy}{{\bm{y}}}
\newcommand{\bmz}{{\bm{z}}}
\newcommand{\bmalpha}{{\bm{\alpha}}}
\newcommand{\bmbeta}{{\bm{\beta}}}
\newcommand{\bmepsilon}{{\bm{\varepsilon}}}

\newcommand{\bmmu}{{\bm{\mu}}}
\newcommand{\bmsigma}{{\bm{\sigma}}}
\newcommand{\bmchi}{{\bm{\chi}}}

\newcommand{\amode}{\bm{\hat{\alpha}}}   % (Kx1) parameter vector at mode
\newcommand{\betamode}{{\bm{\hat{\beta}}}}

\newcommand{\tr}{{\scriptscriptstyle\sf T}}  % transpose

\newcommand{\hypo}   {{\mathcal{H}}}  % Hypothesis
\newcommand{\sspace}{{\mathcal{A}}}  % Outcome subspace (use with subscript)
\newcommand{\cond}{\,|\,}
\newcommand{\normaldist}{{\mathcal N}}
\newcommand{\var}{\mathrm{var}}

\newcommand{\drm}{\mathrm{d}}
\newcommand{\bfac}{\mathrm{BF}}
\newcommand{\hbic}{\hypo_{\textsc{bic}}}
\newcommand{\haic}{\hypo_{\textsc{aic}}}

\newcommand{\Ksim}{{S}}
\newcommand{\smsim}{{\scriptscriptstyle S}}
\newcommand{\struta}{\mbox{\rule[0pt]{0pt}{9pt}}} % strut to make square root readable
\begin{document}

\begin{center}
  \textbf{\Large Bayesian Model Selection for Misspecified Models in Linear Regression}\\[8pt]
  M.B.\ de Kock${}^1$ and H.C.\ Eggers${}^{1,2}$\\[8pt]
  ${}^1$\textit{Institute of Theoretical Physics and Department of
    Physics, Stellenbosch University, P/Bag X1,}\\
  \textit{7602 Matieland, Stellenbosch, South Africa}\\
  ${}^2$\textit{National Institute for Theoretical Physics,}\\
  \textit{P/Bag X1, 7602 Matieland, South Africa}
\end{center}

\begin{abstract}
  While the Bayesian Information Criterion (\BIC) and Akaike
  Information Criterion (\AIC) are powerful tools for model selection
  in linear regression, they are built on different prior assumptions
  and thereby apply to different data generation scenarios. We show
  that in the finite dimensional case their respective assumptions can 
  be unified within an augmented model-plus-noise space and construct 
  a prior in this space which inherits the beneficial properties of both 
  \AIC\ and \BIC. This allows us to adapt the \BIC\ to be robust against 
  misspecified models where the signal to noise ratio is low.
\end{abstract}

\section{Introduction}
\label{sec:intro}
The selection of a model between multiple competing models is a well 
established tool of data analysis in a wide spectrum of fields ranging 
from ecology to psychology\cite{burnham2003model,vrieze2012model,aho2014model}.
If the true model is one of the candidates then Bayesian methods are consistent 
in that they will select the true model with probability one as the sample size 
increases\cite{Nishii}. On the other hand if the underlying data generating process 
is nonparametric then to estimate the underlying regression function we require a 
minimax-rate optimal rule\cite{shao}. This dichotomy is closely related to the 
competition between \AIC\ and \BIC\ , where \BIC\ represents the Bayesian methods and \AIC\ 
the loss-optimal rules. In general, it is though to be impossible to combine the 
properties of both, see \cite{yang2005can}. 

Our goal is not to address the difference between the parametric and nonparametric
cases but to give a Bayesian construction for linear regression that is robust to 
model misspecification. It is similar to \cite{muller2013}, but does not introduce an 
artificial posterior but augments the likelihood with a larger parameter space. This 
extends the idea of Akaike\cite{akaike1978bayesian} which appeared after 
Akaike \cite{akaike1974new} and Schwartz introduced,\cite{schwarz1978estimating}, 
their information criterion, \AIC\ and \BIC, respectively. Akaike, in the limited 
context of linear regression, considers the model selection problem in a parameter 
space expanded from the $K$-dimensional model space into a larger one and then 
shows how different Bayesian priors in this space give arise to the two different 
information criterion. This reveals that the BIC implicitly includes a prior which 
fixes the extra non-model ``noise parameters'' to be exactly zero, while the \AIC\ 
allows them to vary to some degree around zero. This confirms our own experience in
that as the noise-to-signal ratios is lowered to unity the \BIC\ fails completely.
Statistically speaking, the \BIC\ assumes one of the candidate models is the true 
model while \AIC\ does not. It is this property that we wish to extend to the \BIC\ case.

Armed with this insight, we shall construct priors not only for the
$K$ model parameters, but for an additional set of $L = N-K$ noise
parameters in the spirit of Akaike. Unlike their \BIC\ and \AIC\ 
predecessors, however, these new priors will take into account the
crucial information that the modes of signal parameter priors should
be located some distance away from the origin of parameter space. 
This allows us to adapt the Bayesian Information Criterion to 
not assume that one of the candidate models are true. 

The paper is organised as follows. In Section \ref{sec:blr}, we review
the framework of Bayesian model comparison and linear regression,
augmenting in Section \ref{sec:nsp} the model space by a noise space
in preparation for the new priors. Reconsideration in Section
\ref{sec:prcn} of the priors underlying the \BIC\ and \AIC\ forms the
basis and motivation for the construction of spherically symmetric
priors for both model and noise parameters in Section
\ref{sec:nic}. The results are tested numerically and compared to the
traditional information criteria in Section \ref{sec:smm}, followed by
a brief summary and discussion in Section \ref{sec:dsc}.

\section{Bayesian linear regression}
\label{sec:blr}

\subsection{Bayesian model selection}
\label{sec:bmc}

Given data $\mathcal{D}$, Bayesian model selection is based on the
evidence or marginal likelihood for the model $\hypo_\smK$ which has
$K$ parameters $\bmalpha_\smK = (\alpha_1,\ldots,\alpha_\smK)$. The
evidence is an average over the likelihood
$p(\mathcal{D}\cond \bmalpha_\smK,\hypo_\smK)$ weighted by the
parameter prior $p(\bmalpha_\smK|\hypo_\smK)$,
\begin{align}	
  \label{bsb}
  p(\mathcal{D} \cond \hypo_\smK) 
  &= \int p(\mathcal{D}\cond \bmalpha_\smK, \hypo_\smK)
    \,p(\bmalpha_\smK|\hypo_\smK)\, \drm \bmalpha_\smK.
\end{align}
where $p(\bmalpha_\smK|\hypo_\smK)$ may contain hyperparameters as
necessary.  Bayes' theorem used twice for competing models
$\hypo_\smK$, $\hypo_{\smK'}$ relates the ratio of model posteriors
$p(\hypo_\smK\cond\mathcal{D})$ to the corresponding model evidences
by
\begin{align}
  \label{bsc}
  \frac{p(\hypo_\smK\cond\mathcal{D})}{p(\hypo_{\smK'}\cond\mathcal{D})}
  &= 
    \frac{p(\hypo_\smK)}{p(\hypo_\smK')}\,
    \frac{p(\mathcal{D}\cond \hypo_\smK)}{p(\mathcal{D}\cond\hypo_{\smK'})}.
\end{align}
Barring good reasons to deviate from the Principle of Indifference,
model priors would normally be set equal,
$p(\hypo_\smK) = p(\hypo_\smK') = \tfrac{1}{2}$, in which case the
posterior odds equals the Bayes Factor \cite{kass1995bayes}
\begin{align}
  \label{bsd}
  \frac{p(\hypo_\smK\cond\mathcal{D})}{p(\hypo_{\smK'}\cond\mathcal{D})}
  &= \frac{p(\mathcal{D}\cond \hypo_\smK)}{p(\mathcal{D}\cond\hypo_{\smK'})}
   \equiv \bfac[K,K'].
\end{align}
When more than two models are to be compared, it is convenient to
define a reference model against which all others are measured. In
this paper, we use as reference model $\hypo_\smN$, the case where $N$
data points are modelled by $K=N$ free parameters, implying of course
an exact fit and no noise. Among the $N$ competing models with
$K=1,2,\ldots N$ parameters, the best model is the one with maximal
Bayes Factor or equivalently minimum information criterion
$\mathrm{IC} = -2\log\bfac$.

\subsection{Linear Regression}
\label{sec:lrg}

\begin{figure}
  \centering
  \begin{tikzpicture}
    \draw [<->] (0,6) node (yaxis) [above] {$e_2$}
    |- (10,0) node (xaxis) [right] {$e_1$};
    \draw [->] (0,0)--(-2,2) node (zaxis) [above] {$e_3$};
    \draw[->,thick] (0,0) -- (4,5) node [pos=1,above] {$\bm{z} = \mbox{data}$};
    \draw[->,thick] (5.2,2.6) -- (4,5) node [pos=0.6,right] {$\hat{\bmchi}$};
    \draw[->,thick] (0,0) -- (5.2,2.6) node [pos=0.95,below] {$\hat{\bmf}_\smK$};
    \draw[->,thick] (3.5,1.75) -- (4,5) node [pos=0.5,left] {$\bmchi$};
    \draw[->,thick] (0,0)--(3.5,1.75) node [pos=0.65,above] {$\bmf_\smK$};
    % % % angle
    \draw (3.55,2.025) arc (90:190:0.35) node[pos=0.6,above] {$\theta$};
    % % % Orthogonal
    \draw[thick](5,2.5) -- (4.9,2.7);
    \draw[thick](4.9,2.7) -- (5.1,2.8);
    % % % Draw Ak plane
    \foreach \x in {1,...,8}
    \draw[dashed] (\x-0.866,\x/2+0.5) -- (\x+0.866,\x/2-0.5);
    \draw(0,0) -- (8,4);
    \draw[dashed] (1-0.866,1) -- (8-0.866,4.5);
    \draw[dashed] (1+0.866,0) -- (8+0.866,3.5) node [pos=0.5,below] {$\sspace_\smK$};
    % % % Draw Al plane
    \draw [<->] (9.2,1.6) -- (6.8,6.4) node [pos=1.0,above] {$\sspace_\smL$}; 
    % % % Orthogonal
    \draw(7.9,3.7)--(7.7,4.1);
    \draw(7.7,4.1)--(7.9,4.2);
    \draw(7.9,3.7)--(8.1,3.8);
  \end{tikzpicture}
  \caption{Cartoon of a three-dimensional data space $\sspace_\smN$
    spanned by $(e_1,e_2,e_3)$ partitioned into a two-dimensional
    model space $\sspace_\smK$ and a one-dimensional noise space
    $\sspace_\smL$. While the signal vector $\bmf_\smK$ is always in model
    space, the noise vector $\bmchi$ is contained in noise space
    $\sspace_\smL$ only for the best-fit case $\hat{\bmchi}$ whose length
    is the minimum distance between data vector $\bmz$ and model space.
    \label{fig:dspace}}
\end{figure}
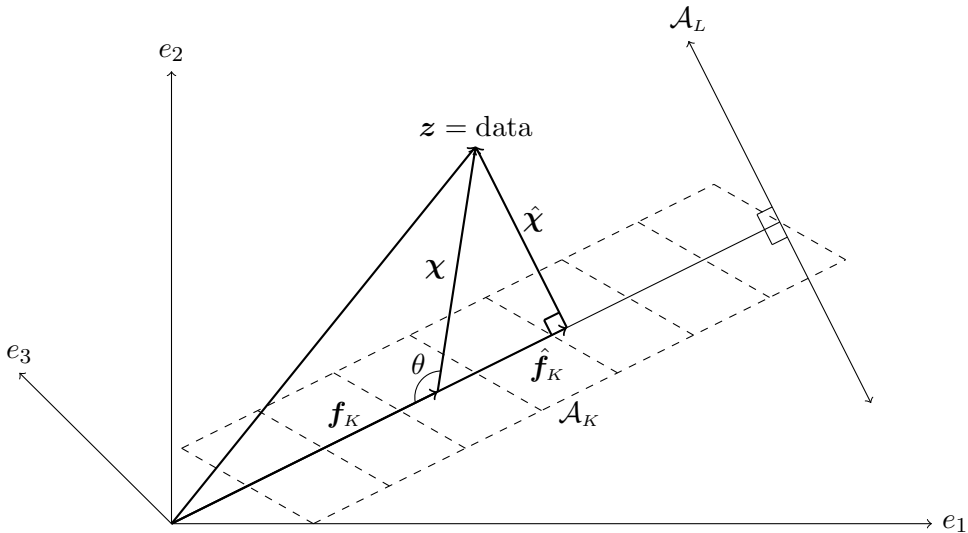

We briefly review the canonical formalism for linear regression and
introduce the language and notation to be used in later sections.
By assumption the data $\mathcal{D}$ comes in the form of $N$ data
points $y_n \in \bmy = (y_1,\ldots,y_\smN)$ measured at locations
$x_n \in \bmx = (x_1,\ldots,x_\smN)$ with fixed experimental
uncertainties $\sigma_n \in \bmsigma$. %
The immediate aim is to find joint distributions (posterior, evidence
etc) of the $K$ coefficients
$\alpha_k \in \bmalpha_\smK = (\alpha_1,\ldots,\alpha_\smK)$ of a
linear model function
\begin{align}
  \label{lry}
  y(x\cond \bmalpha_\smK,K) &= \sum_{k=1}^K f_k(x)\,\alpha_k,
\end{align}
where the choice of basis functions $f_k(x), k=1,\ldots, K$ forms part
of the model specification and we subscript model-dependent quantities
by $K$ in preparation for the extensions of Section \ref{sec:nsp}. By
assumption, the differences
$\varepsilon_n \equiv y_n - y(x_n\cond\bmalpha_\smK,K)$ between each
data point and the corresponding model point are normally
distributed, %
\begin{align}
  \label{bmep}
  p(\bmepsilon|\bmsigma) 
  &= \prod_{n=1}^N \frac{e^{-\varepsilon_n^2/2\sigma_n^2} }{\sigma_n\sqrt{2\pi}},
\end{align}
resulting in a joint likelihood
\begin{align}
  \label{likz}
  L\left[\bmalpha_\smK\right] 
  = p(\bmy\cond\bmalpha_\smK,K) 
  &= \prod_{n=1}^N \frac{1}{\sigma_n\sqrt{2\pi}}
    \exp\left[-\frac{1}{2}\left(
    \frac{y_n}{\sigma_n} - \sum_{k=1}^K \frac{f_k(x_n)}{\sigma_n}\alpha_k\right)^2\,\right].
\end{align}
The $K$-dimensional model space $\sspace_\smK$ is spanned by
$N$-dimensional basis vectors
\begin{align}
  \bmv_k^\tr &= [f_k(x_1)/\sigma_1,\ldots,f_k(x_\smN)/\sigma_\smN]\qquad k = 1,\ldots,K,
\end{align}
which together constitute the $(N{\times}K)$-dimensioned design matrix
%\begin{align}
  $\matX_\smK 
  = [\bmv_1,\ldots,\bmv_\smK]$.
 % = \begin{bmatrix} f_1[x_1]/\sigma_1 & f_2[x_1]/\sigma_1 & \cdots & f_\smK[x_1]/\sigma_1 \\
 %   f_1[x_2]/\sigma_2 & f_2[x_2]/\sigma_2 & \cdots & f_\smK[x_2]/\sigma_2 \\
 %   \vdots 	 & \vdots 	& \ddots & \vdots \\
 %   f_1[x_\smN]/\sigma_\smN & f_2[x_\smN]/\sigma_\smN & \cdots & f_\smK[x_\smN]/\sigma_\smN 
 % \end{bmatrix}.
%\end{align}
In terms of the standardised data vector
$\bmz=[y_1/\sigma_1,\ldots,y_\smN/\sigma_\smN]$ and collecting
constants into $C = (2\pi)^{-N/2}[\textstyle\prod_n\sigma_n]^{-1}$,
the likelihood can be written in three ways,
\begin{align}
  \label{lika}
  L\left[\bmalpha_\smK\right] 
  &= C \exp\left[-\tfrac{1}{2}
    \left(\bmz{-}\matX_\smK\bmalpha_\smK\right)^\tr\!
    \left(\bmz{-}\matX_\smK\bmalpha_\smK\right)\right]
  \\
  &= C \exp\left[-\tfrac{1}{2} (\bmz {-} \bmf_\smK)^\tr (\bmz {-} \bmf_\smK) \right]
  \ =\ C \exp\left[-\tfrac{1}{2} \bmchi^\tr\bmchi \right]
\end{align}
where $\bmf_\smK = \matX_\smK\bmalpha_\smK = \sum_{k=1}^K \bmv_k\alpha_k$
is the model-dependent vector aspiring to approximate the data vector
$\bmz$ and $\bmchi = \bmz - \bmf_\smK$ is the discrepancy between data and
model. Description of the data $\bmz$ is thereby decomposed into a
``noise'' component $\bmchi$ and a ``signal'' component $\bmf_\smK$. As
illustrated in Figure~\ref{fig:dspace}, $\bmchi$ and $\bmf_\smK$ are in
general not orthogonal. The length of the minimum-chisquared vector
$\hat{\bmchi}$ represents the minimum distance between the data vector $\bmz$
and model space $\sspace_\smK$, so that it is orthogonal to model
space $\bmf_\smK^\tr\hat{\bmchi} = 0$ for all $\bmf_\smK^\tr$. The resulting maximum-signal
vector can be found directly from
\begin{align}
  \hat{\bmf}_\smK &= \matX_\smK\amode_\smK = \textstyle\sum_{k=1}^K \bmv_k\amode_k
\end{align}
where the maximum-likelihood parameter vector is determined by the
usual Moore-Penrose inverse
\begin{align}
  \label{ptg}
  \amode_\smK &= \matH_\smK^{-1}\matX_\smK^\tr \bmz,
\end{align}
with $\matH_\smK = \matX_\smK^\tr\matX_\smK$ the Hessian with elements
$(\matH)_{k k'} = \bmv_k^\tr\bmv_{k'} = \sum_n
f_k(x_n)f_{k'}(x_n)/\sigma_n^2$.  Since $\bmz = \hat{\bmchi} + \hat{\bmf}_\smK$, the
squared data vector $z^2 = \bmz^\tr\bmz$ can hence be written as the
Pythagorean sum of the usual minimum chisquared $\chi^2 = \hat{\bmchi}^\tr\hat{\bmchi}$ and
the squared signal vector $F_\smK^2 = \hat{\bmf}_\smK^\tr\hat{\bmf}_\smK$,
\begin{align}
  \label{pth}
  z^2 &= \chi^2 + F_\smK^2.
\end{align}
Following the usual diagonalisation by orthonormal eigenvector matrix
$\matS_\smK$ and rescaling by diagonal eigenvalue matrix $\matL_\smK$
and transforming to hyperspherical parameters
$\bmbeta_\smK = \matL_\smK^{\!\!1/2}\matS_\smK^\tr\bmalpha_\smK$, and
corresponding modes
$\betamode_\smK = \matL_\smK^{\!\!1/2}\matS_\smK^\tr\amode_\smK$, the
likelihood becomes
\begin{align}
  \label{likb}
  L\left[\bmalpha_\smK\right] 
  &= C \exp\left[-\tfrac{1}{2}\chi^2 -\tfrac{1}{2}
    (\bmalpha_\smK{-}\amode_\smK)^\tr \matH_\smK
    (\bmalpha_\smK{-}\amode_\smK)\right],
    \\
  \label{likc}
  L\left[\bmbeta_\smK\right] 
  &=  C \exp\left[-\tfrac{1}{2}\chi^2
    -\tfrac{1}{2}
    (\bmbeta_\smK-\betamode_\smK)^\tr 
    (\bmbeta_\smK-\betamode_\smK)\right],
\end{align}
and the squared signal vector transforms to
\begin{align}
  \label{rsq}
  F_\smK^2 
  &
    = \hat{\bmf}_\smK^\tr\hat{\bmf}_\smK
    = \amode_\smK^\tr\matH_\smK\amode_\smK 
    = \betamode_\smK^\tr\betamode_\smK
    = \sum_{k=1}^K \hat{\beta}_k^2.
\end{align}
All of the above is the standard fare of linear regression.

\section{Model space and noise space}
\label{sec:nsp}

The expanded model-noise space introduced in this Section is best
understood in the context of Akaike's rederivation in a Bayesian
framework of the \BIC\ and \AIC\ in \cite{akaike1978bayesian}. His central
message was that both could be understood by introducing, over and
above the $K$ parameters $\bmbeta_\smK = (\beta_1,\ldots,\beta_\smK)$
making up the model, an additional set of parameters
$\bmbeta_\smL = (\beta_{\smK+1},\ldots,\beta_\smM)$ with
$K < M \leq N$, for which particular choices of priors yield the \BIC\ 
and \AIC. Details of the derivation are postponed to Section
\ref{sec:prcn}.

The introduction of additional parameters in Akaike's derivations allows 
the method to account for misspecified model functions in that if there is
some signal left the noise parameters would be able to fit the shift in 
the residuals.
%is
%more than an ad hoc measure.  Along with \cite{akaike1978bayesian} and
%\cite{jeffreys1998theory} we would argue that, since it is impossible
%to know the value of a parameter with infinite accuracy even in the
%presence of data, setting a parameter to exactly zero in a prior,
%before any data is taken, is even harder to defend.  The very concept
%of a parameter is implicitly based on uncertainty, and the mere fact
%that model comparison is undertaken implies uncertainty regarding both
%the particular model and its parameter values.
% 
The difference between a model $\hypo_\smK$ with $K$ parameters and
another model $\hypo_{\smK+1}$ with $K{+}1$ parameters must then be
found not in the existence or nonexistence of additional parameter
$\beta_{\smK+1}$ but in different priors
$p(\beta_{\smK+1}|\hypo_\smK)\neq p(\beta_{\smK+1}|\hypo_{\smK+1})$.
In this view, all \textit{model parameters} $\bmbeta_\smK$ should be
assigned priors which allow them to exhibit large deviations from
zero, while the additional \textit{noise parameters} $\bmbeta_\smL$
should be assigned priors which are not exactly zero but restricted to
small intervals around the origin.

Taking this line of thought to its logical conclusion, we let
$M\equiv N$ and introduce $L=N-K$ additional noise parameters
$\beta_\ell \in \bmbeta_\smL = (\beta_{\smK+1},\ldots,\beta_\smN)$
along with $L$ additional basis functions
$\{f_\ell(x)\}_{\ell=K{+}1}^N$ spanning what we shall call the
\textit{noise space} $\sspace_\smL$.
While the mathematics does not preclude overlap, it seems natural to
demand that model space (also called signal space) $\sspace_\smK$ and
noise space $\sspace_\smL$ partition the data space,
\begin{align}
  \sspace_\smN &= \sspace_\smK\cup\sspace_\smL,
  \qquad \sspace_\smK\cap\sspace_\smL = \emptyset.
\end{align}
In this view, model construction is seen as a successive decomposition
of the data space $\sspace_\smN$ into sequences of partitions
$\{\sspace_\smK,\sspace_\smL\}_{K=1}^N$ with progressively increasing
$K$ and decreasing $L$, with model selection based on the maximum
evidence or Bayes Factor as a function of $K$. %
The partitioning property can be enforced by constructing, if
necessary by a Gram-Schmidt procedure, a set of \textit{noise
  functions} $f_\ell(x)$ which are orthogonal to all model functions
$f_k(x)$,
\begin{align}
  \sum_{n=1}^N f_k(x_n)\,f_\ell(x_n) &= 0
  \qquad \forall\; k=1,\ldots,K \text{ and } \ell=K{+}1,\ldots,N,
\end{align}
thereby ensuring\footnote{The simplest way to ensure block-diagonality
  is to construct a complete orthogonal basis for all $K{=}N$
  functions $f_k$ which would trivially fulfil these requirements. The
  block-diagonal form is, however, more widely applicable.} %
that the Hessian of the complete basis set $\{f_k\}_{k=1}^N$ is
block-diagonal, $\matH_\smN = \matH_\smK{\oplus}\matH_\smL$.
%%%%%%
We note that the basis functions $f_\ell(x)$ may have to be adapted as
$K$ changes to safeguard block-diagonality. The resulting sequence of
models is therefore not nested in the strict sense.

As already mentioned in Section \ref{sec:lrg}, $\bmf_\smK^\tr\hat{\bmchi} = 0 $ for all
$\bmf_\smK$ because, with the help of Eq.~(\ref{ptg}),
$\bmalpha_\smK^\tr \matX_\smK^\tr(\bmz - \matX_\smK\amode_\smK) =
\bmalpha_\smK^\tr \matX_\smK^\tr(\bmz -
\matX_\smK\matH_\smK^{-1}\matX_\smK^\tr\bmz) = 0$. %
Together with $\sspace_K\cap\sspace_L=\emptyset$, this means that that
$\hat{\bmchi}$ is a vector in $\sspace_\smL$ and hence has a representation in
the noise-space basis
$\bmv_\ell = [f_\ell(x_1)/\sigma_1,\ldots,f_\ell(x_\smN)/\sigma_\smN]$
with coefficients
$\amode_\smL = (\hat{\alpha}_{\smK+1},\ldots,\hat{\alpha}_\smN)$ or
equivalently in terms of the noise-space design matrix
$\matX_\smL = [\bmv_{\smK{+}1},\ldots,\bmv_\smN]$,
\begin{align}
  \hat{\bmchi} &= \sum_{\ell=\smK+1}^N \bmv_\ell\,\hat{\alpha}_\ell
         = \matX_\smL \amode_\smL.
\end{align}
Diagonalisation by $\matS_\smL$ and rescaling by $\matL_\smL$ in noise
space results in
$\hat{\bmchi} = \matX_\smL \matS_\smL\matL_\smL^{-1/2}\betamode_\smL$ with
$\betamode_\smL = \matL_\smL^{1/2}\matS_\smL^\tr\amode_\smL$ just as
in model space, so that, in close analogy with Eq.~(\ref{rsq}),
\begin{align}
  \label{qkb}
  \chi^2 
  &
    = \hat{\bmchi}^\tr\hat{\bmchi} 
    = \amode_\smL^\tr \matH_\smL \amode_\smL
    = \betamode_\smL^\tr \betamode_\smL
    = \sum_{\ell=K+1}^{N} \hat{\beta}^2_\ell.
\end{align}
Apart from the requirement that the basis functions $f_\ell(x)$ must
be orthogonal to those in model space, their specific choice is
arbitrary; correspondingly, individual coefficients $\bmalpha_\smL$,
$\bmbeta_\smL$ and their maximum-likelihood cases $\amode_\smL$ and
$\betamode_\smL$ are not fixed by the model. All that matters is that
$\chi^2$ can be written as a sum of $L=N{-}K$ squared components
$\hat{\beta}_\ell$.

The extension from $\sspace_\smK$ to the larger space
$\sspace_\smK\cup\sspace_\smL$ has an important consequence for
the likelihood and evidence.  Rather than using the conventional
$K$-dimensional version which would result from Eq.~(\ref{likb}),
\begin{align}
  \label{likd}
  L[\bmbeta_\smK]
  &= p(\bmy\cond \bmbeta_\smK,\hypo)
  =  C \exp\left[-\tfrac{1}{2}\betamode_\smL^\tr\betamode_\smL
    -\tfrac{1}{2}
    (\bmbeta_\smK-\betamode_\smK)^\tr 
    (\bmbeta_\smK-\betamode_\smK)\right],
\end{align}
the limited model $\sum_{k=1}^K f_k(x)\alpha_k$ in the likelihood of
Eq.~(\ref{likz}) is replaced with the full set,
\begin{align}
  \label{likf}
  L\left[\bmalpha_\smK,\bmalpha_\smL\right] 
  &= C \exp\left[-\frac{1}{2}\sum_{n=1}^N\left(
    \frac{y_n}{\sigma_n} 
    - \sum_{k=1}^K     \frac{f_k(x_n)}{\sigma_n}\alpha_k
    - \sum_{\ell=K+1}^N \frac{f_\ell(x_n)}{\sigma_n}\alpha_\ell\right)^2
    \,\right],
\end{align}
which for block-diagonal $\matH_\smN=\matH_\smK\oplus\matH_\smL$ takes
the form
\begin{align}
  \label{likg}
  p(\bmy\cond\bmbeta_\smK,\bmbeta_\smL)
  = 
  L[\bmbeta_\smK,\bmbeta_\smL]
  &= C \exp\left[
    -\tfrac{1}{2}
    (\bmbeta_\smL-\betamode_\smL)^\tr 
    (\bmbeta_\smL-\betamode_\smL)
    -\tfrac{1}{2}
    (\bmbeta_\smK-\betamode_\smK)^\tr 
    (\bmbeta_\smK-\betamode_\smK)\right],
\end{align}
for which the evidence factorises into noise and signal parts,
\begin{align}
  \label{evxt}
  p(\bmy\cond K,\hypo) 
  &= \int \drm\bmbeta_\smK\,\drm\bmbeta_\smL\, L[\bmbeta_\smK,\bmbeta_{\smL}]\,
    p(\bmbeta_\smK|\hypo_\smK)\,p(\bmbeta_{\smL}|\hypo_\smL)
    \nonumber\\
  &= C
     \left\{\int \drm \bmbeta_\smL\,
     \exp\left[-\tfrac{1}{2} (\bmbeta_\smL-\betamode_\smL)^\tr (\bmbeta_\smL-\betamode_\smL)\right]
     \,p(\bmbeta_\smL|\hypo_\smL) \right\}
    \nonumber\\
   &\quad\times  
    \left\{\int \drm \bmbeta_\smK\,
    \exp\left[-\tfrac{1}{2} 
      (\bmbeta_\smK-\betamode_\smK)^\tr (\bmbeta_\smK-\betamode_\smK)\right]
    \,p(\bmbeta_\smK|\hypo_\smK)
    \right\}.
\end{align}
We briefly consider the case $K{=}N, L{=}0$ as this constitutes our
reference model for Bayes Factors. The design matrix $\matX_\smN$ is
invertible so that $\amode_\smN = \matX_\smN^{-1}\bmz$, all the data
becomes signal ($\bmz = \hat{\bmf}_{\scriptscriptstyle K=N}$,
$z^2 = F_\smN^2$), there is no noise
($\hat{\bmchi}_{\scriptscriptstyle{L} =0}=0$), and the model amounts to a
change of basis for $\sspace_\smN$ from $\bmv_\smN$ to
$\bmalpha_\smN$.

\section{Akaike's BIC/AIC priors in model and noise space}
\label{sec:prcn}

We now rederive Akaike's central insight in the language of
model and noise space. 
The \BIC\ results when the priors for the model parameters
$\beta_k,k=1,\ldots,K$ are normally distributed with variance $N$,
while the additional parameters $\beta_\ell, \ell = K{+}1,\ldots,N$
are set to zero exactly by means of Dirac delta functions,\footnote{ %
The reasoning behind setting the model prior variances to $N$ is
that the exponent of the likelihood $L[\bmalpha_\smK,\bmalpha_\smL]$
scales roughly with $N$ as long as parameter-parameter correlations
do not dominate, so that $\bmbeta \approx \bmalpha/\sqrt{N}$ and
$\bmbeta^\tr\bmbeta \approx 1/N$.}
\begin{align}
  \label{ccb}
  & p(\beta_k|\hbic) = \frac{1}{\sqrt{2\pi N}} e^{-\beta_k^2/2N},
  && p(\beta_\ell|\hbic) = \delta(\beta_{\ell}).
\end{align}
The evidence (\ref{evxt}) and corresponding $K{=}N$ reference evidence
are then
\begin{align}
  \label{ccc}
  p(\bmy\cond K,\hbic) 
  &= C (N{+}1)^{-K/2} 
  \exp\left[-\frac{\chi^2}{2} - \frac{F_\smK^2}{2(N{+}1)} \right],
    \\
  \label{cce}
  p(\bmy\cond N,\hbic) 
  &= C (N{+}1)^{-N/2} 
  \exp\left[- \frac{\chi^2 + F_\smK^2}{2(N{+}1)} \right],
\end{align}
yielding a Bayes Factor
\begin{align}
  \label{ccf}
  \bfac[K,N] 
  &= (N{+}1)^{(N-K)/2} \exp\left[-\frac{N \chi^2}{2(N{+}1)} \right].
\end{align}
Dropping $K$-independent constants and assuming $N\gg 1$, we recover
the BIC from the logarithm
\begin{align}
  \label{ccg}
  -2\log \bfac[K,N] &\ \simeq\  \chi^2 + K\log N \ =\ \mbox{\BIC}
\end{align}
since $\chi^2 \propto -2\log$(maximum likelihood). %
In rederiving the \AIC, \cite{akaike1978bayesian} similarly suggested
that model and noise parameters be treated on the same basis but with
different scales for their priors,
\begin{align}
  \label{cch}
  & p(\beta_k|\haic) = \frac{1}{\sqrt{2\pi \Delta^2}} e^{-\beta_k^2/2\Delta^2},
 && p(\beta_\ell|\haic) = \frac{1}{\sqrt{2\pi \delta^2}} e^{-\beta_\ell^2/2 \delta^2}
\end{align}
resulting in evidence, reference evidence and Bayes Factor
\begin{align}
  \label{cci}
  p(\bmy\cond K,\haic) 
  &= \frac{C}
    {\left[1{+}\Delta^2\right]^{K/2} \left[1{+}\delta^2\right]^{L/2}} 
    \exp\left[-\frac{\chi^2}{2(1{+}\delta^2)} - \frac{F_\smK^2}{2(1{+}\Delta^2)} \right],
    \\
  \label{ccj}
  p(\bmy\cond N,\haic) 
  &= \frac{C}
    {\left[1{+}\Delta^2\right]^{N/2} }
    \exp\left[-\frac{\chi^2 {+} F_\smK^2}{2(1{+}\Delta^2)} \right],
    \\
  BF[K,N] 
  &= \left[\frac{1{+}\Delta^2}{1{+}\delta^2}\right]^{L/2} 
    \exp\left[-\left(\frac{1}{1{+}\delta^2} -\frac{1}{1{+}\Delta^2} \right)\frac{\chi^2}{2} \right],
\end{align}
and information criterion
\begin{align}
  \label{cck}
  -2 \log BF[K,N] 
  &= \left(\frac{1}{1{+}\delta^2} -\frac{1}{1{+}\Delta^2} \right) 
    \chi^2 -  \left(N{-}K\right) \log \frac{1{+}\Delta^2}{1{+}\delta^2}.
\end{align}
At this point, Akaike argued that $\delta$ and $\Delta$ should
approach 1 from above and below, where 1 represents the situation of
equal signal and noise magnitude. This is the critical case where it is difficult to distinguish between model and noise and the Bayes Factor will tend to zero. To find the next to leading order behaviour of the Bayes Factor we as Akaike take the limit
$\delta\rightarrow 1^{-}$ and $\Delta\rightarrow 1^{+}$, and recover the \AIC, %
\begin{align}
  \label{ccl}
  \lim_{\substack{\delta\rightarrow 1\\\Delta\rightarrow 1}}
  \left(\frac{1}{1+\delta^2} -\frac{1}{1+\Delta^2} \right)^{\!\!-1}
  \!\!\left( -2\log BF[K,N]\right) 
  &\ =\ \chi^2 + 2K \ =\ \mathrm{\AIC}.
\end{align}
%%%%%%%%%%%%%%%%%%%%%%%%%%%%%%%%%%%%%%%%%%%%%%%%%%%%%%%%%%%%%%%%%
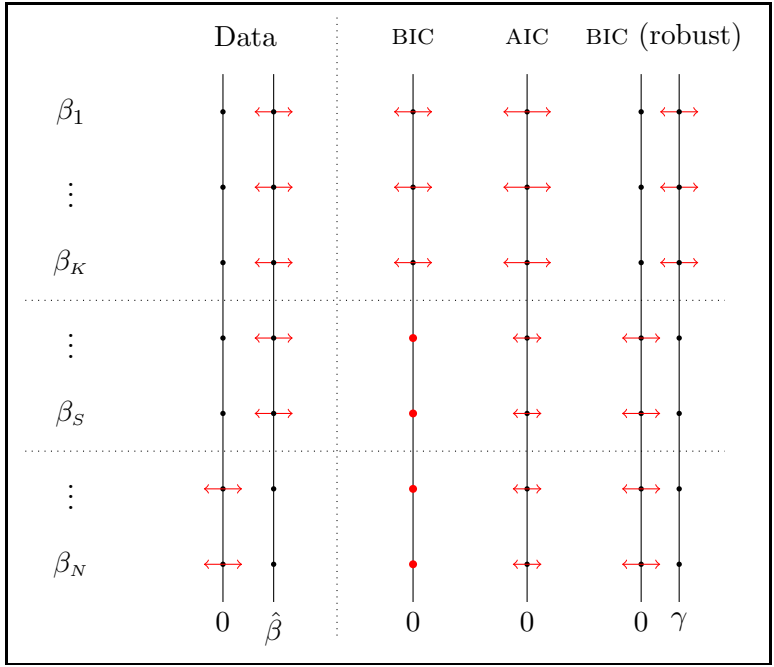
\begin{figure}
  \centering
  \fbox{
    \begin{tikzpicture}
      % % Dots
      \foreach \x in {2,4.5,6,7.5} {
        \foreach \y in {0,1,2,3,4,5,6} {
          \fill[black] (\x,-\y) circle (1pt);
        }
        \draw (\x,1/2) -- (\x,-6.5) node[anchor=north] {0};
      }
      % % Labels
      \node at (0,0) {$\beta_1$};
      \node at (0,-1) {$\vdots$};
      \node at (0,-2) {$\beta_\smK$};
      \node at (0,-3) {$\vdots$};
      \node at (0,-4) {$\beta_{\smS}$};
      \node at (0,-5) {$\vdots$};
      \node at (0,-6) {$\beta_\smN$};
      \node at (2.3,1) {Data};
      \node at (4.5,1) {\BIC};
      \node at (6,1) {\AIC};
      \node at (7.8,1) {\BIC\ \mbox{(robust)}};
      % % Data
      \draw (2+2/3,1/2) -- (2+2/3,-6.5) node[anchor=north] {$\hat{\beta}$};
      \foreach \y in {0,1,2,3,4} {
        \draw[red,->] (2+2/3,-\y)--(2-1/4+2/3,-\y);
        \draw[red,->] (2+2/3,-\y)--(2+1/4+2/3,-\y);
        \fill (2+2/3,-\y) circle (1pt);
      }
      \foreach \y in {5,6} {
        \draw[red,->] (2,-\y)--(2-1/4,-\y);
        \draw[red,->] (2,-\y)--(2+1/4,-\y);
        \fill (2+2/3,-\y) circle (1pt);
      }
      % % BIC
      \foreach \y in {0,1,2} {
        \draw[red,->] (4.5,-\y)--(4.5-1/4,-\y);
        \draw[red,->] (4.5,-\y)--(4.5+1/4,-\y);
      }
      \foreach \y in {3,4,5,6} {
        \fill[red] (4.5,-\y) circle (1.5pt);
      }
      % % AIC
      \foreach \y in {0,1,2} {
        \draw[red,->] (6,-\y)--(6-5/16,-\y);
        \draw[red,->] (6,-\y)--(6+5/16,-\y);
      }
      \foreach \y in {3,4,5,6} {
        \draw[red,->] (6,-\y)--(6-3/16,-\y);
        \draw[red,->] (6,-\y)--(6+3/16,-\y);
      }
      % % BICc
      \draw (8,1/2) -- (8,-6.5) node[anchor=north] {$\gamma$};
      \foreach \y in {0,1,2} {
        \draw[red,->] (8,-\y)--(8-1/4,-\y);
        \draw[red,->] (8,-\y)--(8+1/4,-\y);
        \fill (8,-\y) circle (1pt);
      }
      \foreach \y in {3,4,5,6} {
        \draw[red,->] (7.5,-\y)--(7.5-1/4,-\y);
        \draw[red,->] (7.5,-\y)--(7.5+1/4,-\y);
        \fill (8,-\y) circle (1pt);
      }  
      % % Lines
      \draw[dotted] (3.5,1.25) -- (3.5,-7);
      \draw[dotted] (-0.6,-2.5) -- (8.6,-2.5);
      \draw[dotted] (-0.6,-4.5) -- (8.6,-4.5);
    \end{tikzpicture} 
  }
  \caption{%
    Generic regions where model and noise parameters for the data with
    $\Ksim$ true parameters and the information criteria with $K$
    model parameters are significantly nonzero.  The inconsistency of
    the \BIC\ and \AIC\ cases \textit{vis \`a vis} the data is removed by
    the robust \BIC\ derived in Section \ref{sec:nic}.}
  \label{fig:schema}
\end{figure}
%%%%%%%%%%%%%%%%%%%%%%%%%%%%%%%%%%%%%%%%%%%%%%%%%%%%%%%%%%%%%%%%%
Figure~\ref{fig:schema} summarises schematically the generic form of
the data and model and noise parameter priors for the various
information criteria.  If the true behaviour of the system resulted
from some true model with $\Ksim$ parameters plus noise, the data
would correspond to $\Ksim$ non-zero parameters $\hat{\beta}_k$ and
$N{-}\Ksim$ near-null parameters, both within some uncertainty range;
this is sketched in the two leftmost columns.  The third and fourth
columns in Fig.~\ref{fig:schema} remind us that \BIC\ and \AIC\ set priors
for both model and noise parameters centered around zero, differing
only in the scale of the variation around zero for the noise
parameters. The \BIC\ conflates probabilistic intervals for model
parameters with point probabilities for noise parameters, a strong
assumption which reduces its effectiveness for weak-signal
cases. While consistently using intervals for all parameters, the \AIC\
fails to take account of the fact that model parameters will usually
not be centered around zero.  Consistent with the generic data
behaviour, the robust version of \BIC\ displayed in the last two columns explicitly shifts
model parameter priors away from zero with the help of a
hyperparameter $\gamma$ and consistently uses intervals for all.

\section{Noncentral radial priors and information criterion}
\label{sec:nic}

The lesson of Fig.~\ref{fig:schema} is that properties and performance
of information criteria depend crucially both on their treatment of
noise parameters and the location of the model parameters priors'
modes.
Seen from this perspective, a generic weakness of the \AIC\ and \BIC\ is
self-evident: their model parameter priors are maximal near zero,
while the likelihood will be maximal for nonzero values of
$\bmbeta_\smK$, resulting in poor overlap of prior and likelihood.
This is exactly the problem which the Empirical Bayes criterion tries
to correct, albeit in a nonrigorous way \cite{george2000calibration}.

There is a good reason, of course, to maximise these priors around the
origin. By definition, the state of knowledge embodied in a prior
excludes the location of the data-dependent maximum $\betamode_\smK$,
so that, barring supplementary prior information, the origin becomes
the preferred mode for $p(\bmbeta_\smK|\hypo_\smK)$.
However, while $\betamode_\smK$ itself may not be used, we can and
should take into consideration the generic fact that for a good model
the parameters $\betamode_\smK$ will be nonzero. We do not know
\textit{where} in $\sspace_\smK$ the $\betamode_\smK$ is located, but
we do know that the \textit{posterior model radius}
$(\betamode_\smK^\tr\betamode_\smK^{})^{1/2}$ is significantly nonzero
for any and all data, which implies that the prior for what we call
the \textit{prior model radius} ${\parallel}\bmbeta_\smK{\parallel} =
(\bmbeta_\smK^\tr\bmbeta_\smK)^{1/2}$ should be chosen to be
significantly nonzero.
Moreover, the identity $\hat{\bmf}_\smK^\tr\hat{\bmf}_\smK^{} = \betamode_\smK^\tr\betamode_\smK^{}$
in Eq.~(\ref{rsq}) and generally $\bmf_\smK^\tr\bmf_\smK^{} =
\bmbeta_\smK^\tr\bmbeta_\smK^{}$ imply that \textit{the same model
  radius} ${\parallel}\bmbeta_\smK{\parallel}$ sets the scale both in model space
$\sspace_\smK$ and in the corresponding parameter space $\sspace_{\beta,\smK}$.
Likewise, we have generic knowledge that a good model will be
characterised by a near-zero \textit{posterior noise radius}
${\parallel}\betamode_\smL{\parallel} = (\betamode_\smL^\tr\betamode_\smL)^{1/2}$ for
which a near-zero \textit{prior noise radius} ${\parallel}\bmbeta_\smL{\parallel} =
(\bmbeta_\smL^\tr\bmbeta_\smL)^{1/2}$ is of course appropriate, and
\textit{the same noise radius} $q = {\parallel}\bmbeta_\smL{\parallel}$ sets the scale in
both noise space $\sspace_\smL$ and its parameter space
$\sspace_{\beta,\smL}$.
These considerations are compactly summarised in the last column of
Fig.~\ref{fig:schema} and in the equation set
\begin{center}
  \renewcommand{\arraystretch}{1.6}
  \begin{tabular}{r @{\extracolsep{5ex}}  c @{\extracolsep{5ex}} c} 
    & Model & Noise \\
    Prior 
    & ${\parallel}\bmbeta_\smK{\parallel} 
    % A strut to make square roots readable
    = \sqrt{\bmbeta_\smK^\tr\bmbeta_\smK^{}\struta} = \sqrt{\bmf_\smK^\tr\bmf_\smK^{}\struta} > 0 $ 
    & ${\parallel}\bmbeta_\smL{\parallel} 
    = \sqrt{\bmbeta_\smL^\tr\bmbeta_\smL^{}\struta} = \sqrt{\bmchi^\tr\bmchi^{}\struta} \simeq 0$
    \\ 
    Posterior 
    & ${\parallel}\betamode_\smK{\parallel} 
    = \sqrt{\betamode_\smK^\tr\betamode_\smK^{}} = \sqrt{\hat{\bmf}_\smK^\tr\hat{\bmf}_\smK^{}} > 0$
    & ${\parallel}\betamode_\smL{\parallel} 
    = \sqrt{\betamode_\smL^\tr\betamode_\smL^{}} = \sqrt{\hat{\bmchi}^\tr\hat{\bmchi}^{}\struta} \simeq 0$
    \\
  \end{tabular}
  \renewcommand{\arraystretch}{1.0}
\end{center}
All of this constitutes prior knowledge without reference to the
particulars of the data. 
Crucially, this knowledge pertains to the \textit{radii}.  The model
evidence is therefore expanded in terms of two radial parameters $q$
and $r$ and two radial priors,
\begin{align}
  \label{ncc} 
  p(\bmy\cond \hypo_\smK) %
  &= \int_0^\infty \!\drm r\,\drm q \, p(r\cond\hypo_r)\,p(q\cond\hypo_q)%
  \,p(\bmy\cond r,q,\hypo_\smK),
\end{align}
where the evidence conditioned on $r$ and $q$ is
\begin{align}
  \label{ncd} 
  p(\bmy\cond r,q,\hypo_\smK) %
  &= \int \drm\bmbeta_\smK \,\drm\bmbeta_\smL \,
  p(\bmy\cond\bmbeta_\smK,\bmbeta_\smL)
  \,p(\bmbeta_\smK | r)
  \,p(\bmbeta_\smL | q).
\end{align}
Given the factorised likelihood (\ref{likg}), the conditioned
evidence also factorises,
\begin{align}
  \label{nce}
  p(\bmy\cond r,q,\hypo_\smK) 
  &= p(\bmy\cond q,\hypo_\smK) \, p(\bmy\cond r,\hypo_\smK)
  \\
  %\label{ncf}
  &= \int \drm\bmbeta_\smL\, p(\bmbeta_\smL | q)\, 
  e^{-(\bmbeta_\smL-\betamode_\smL)^\tr (\bmbeta_\smL-\betamode_\smL)/2}
  \int \drm\bmbeta_\smK\, p(\bmbeta_\smK | r)\, 
  e^{-(\bmbeta_\smK-\betamode_\smK)^\tr (\bmbeta_\smK-\betamode_\smK)/2}.
    \nonumber
\end{align}
With only radial information available, the prior for $\bmbeta_\smK$
must be uniform on the $K$-hypersphere,
\begin{align}
  \label{ncg}
  p(\bmbeta_\smK\cond r)
  &= \frac{\Gamma(K/2)\,\delta(r - {\parallel}\bmbeta_\smK{\parallel})}
    {2\pi^{K/2}\,r^{K-1}},
\end{align}
with the Dirac delta function constraining the vector $\bmbeta_\smK$
to the surface of the $K$-sphere of radius $r$, while spherical
symmetry on the $L$-hypersphere with radius $q$ requires
\begin{align}
  \label{nch}
  p(\bmbeta_\smL\cond q)
  &= \frac{\Gamma(L/2)\,\delta(q - {\parallel}\bmbeta_\smL{\parallel})}
    {2\pi^{L/2}\,q^{L-1}}.
\end{align}
As shown in \cite{dekock2017bayesian}, the conditioned evidence
(\ref{ncd}) can be expressed in closed form,
\begin{align}
  \label{nci}
  p(\bmy\cond r,q,\hypo_\smK) 
  &= C
    \exp\left[-\frac{1}{2}\left(F_\smK^2 {+} r^2\right)\right]
    {}_0F_1\left[\frac{K}{2}\bigg| \frac{r^2 F_\smK^2}{4} \right] 
    \nonumber\\
  &\quad  \times 
    \exp\left[-\frac{1}{2}\left(\chi^2 {+} q^2\right)\right]
    {}_0F_1\left[\frac{L\,}{2}\bigg| \frac{q^2\chi^2}{4} \right].
\end{align}
We now turn to the radial priors. To capture the generic information
$|\beta_\smL| \simeq 0$, we presuppose that
$\bmbeta_\smL \sim \normaldist(0,\matI_\smL\delta^2)$, a normal
distribution with zero mode and variance $\delta^2$, so that the prior
for the radius $q$ is a chi-squared distribution or Gamma Distribution
in $q^2/2\delta^2$ with hyperparameter $\delta$ and $L = N-K$ as
usual,
\begin{align}
  \label{ncj}
  p(q\cond \delta,L) 
  &= \frac{q}{\delta^2}\left(\frac{q^2}{2\delta^2}\right)^{(L/2)-1}
    \frac{e^{-q^2/2\delta^2}}{\Gamma(L/2)}.
\end{align}
Likewise projecting a $K$-dimensional normal distribution
$p(\bmbeta_\smK|\Delta,\bmmu) =
\normaldist(\bmmu,\matI_\smK\Delta^2)$ with nonzero mode $\bmmu =
(\mu_1,\ldots,\mu_\smK)$ and variance $\mathbb{I}_\smK\Delta^2$ onto
radius $r$ results in a noncentral Gamma Distribution with radial
hyperparameter $\gamma = [\sum_k \mu_k^2]^{1/2}$, %
\begin{align}
  \label{nck}
  p(r\cond \Delta,\gamma,K) 
  &= \frac{r}{\Delta^2}  \left(\frac{r^2}{2\Delta^2}\right)^{(K/2)-1}
    \frac{e^{-(r^2+\gamma^2)/2\Delta^2}}
    {\Gamma(K/2)}
    \;\; {}_0F_1\left[\frac{K}{2}\bigg| \frac{\gamma^2 r^2}{4\Delta^4}\right].
\end{align}
Later, we shall interpret $\gamma$ as a signal-to-noise ratio, and
with $\gamma\to 0$, $p(r\cond \Delta,\gamma,K)$ consistently reverts
to the ordinary Gamma Distribution characteristic of noise. The
noncentral Gamma Distribution results from the projection of 
$\mathcal{N}(\bmmu,\matI_\smK\Delta^2)$ onto the
squared radius $p(r^2| \Delta,\bmmu) = \int \drm\,\bmbeta_\smK
p(r^2|\bmbeta_\smK)\, p(\bmbeta_\smK| \bmmu,\Delta)$ and
using the integral representation of the Dirac delta function %
\begin{align}
  \label{ncl}
  p(r^2\cond \bmbeta_\smK) = 
  \delta(r^2 - \textstyle\sum_k\beta_k^2)
  &= \frac{1}{2\pi i}
  \int_{-i\infty}^{i\infty} \drm s\,\exp[s(r^2 - \textstyle\sum_k\beta_k^2)],
\end{align}
whereby
\begin{align}
  \label{ncm}
  p(r^2\cond \Delta,\gamma)
  &= \frac{1}{(2\pi\Delta^2)^{K/2}}
  \int \drm\bmbeta_\smK\,\delta(r^2 - {\textstyle\sum_k}\beta_k^2)\,
  \exp\left[-\sum_k\frac{(\beta_k-\mu_k)^2}{2\Delta^2} \right] 
  \\
  \label{ncn}
  &= \frac{1}{2\pi i}   
  \int_{-i\infty}^{i\infty} \drm s\,[1 + 2s\Delta^2]^{-K/2}
  \,\exp\left[-\frac{s\gamma^2}{(1{+}2s\Delta^2)}\right],
\end{align}
becomes the Noncentral Gamma Distribution with the help of
\cite{bateman_vol2} % page 15, Eq. (6),
\begin{align}
  \label{nco}
  {}_0F_1[b;z] 
  &= \frac{\Gamma(b)}{2\pi i} 
  \int_{-i\infty}^{+i\infty} \drm u\, u^{-b} \exp\left[u + \frac{z}{u}\right].
\end{align}
Inserting Eqs.~(\ref{nci}), (\ref{ncj}) and (\ref{nck}) into
(\ref{ncc}) and using  \cite{bateman_vol2}
\begin{align}
  \label{ncp}
  \int_0^\infty \frac{e^{-r}r^{a-1}}{\Gamma[a]} 
    \; {}_0F_1[a ; x r] \; {}_0F_1[a ; y r] \; \drm r
    &= e^{x+y} \, {}_0F_1[a; xy],
\end{align}
the evidence is found to be %
\begin{align}
  \label{ncq}
  p(\bmy\cond \delta,\Delta,\gamma,K) %
  &= \frac{C}{
    (1+\delta^2)^{L/2}\,
    (1+\Delta^2)^{K/2}
  }\,
  \exp\left[
    -\frac{\chi^2}{2(1+\delta^2)}
    -\frac{(F_\smK^2\Delta^2 + \gamma^2)}{2\Delta^2}
  \right]
  \nonumber\\
  &\quad\times
  \Psi_{(2)}\left[
  \frac{K}{2}, \frac{K}{2}, \frac{K}{2} 
  \; ;\;
  \frac{\Delta^2 F_\smK^2}{2(1{+}\Delta^2)}
  \; , \;
  \frac{\gamma^2}{2\Delta^2(1{+}\Delta^2)}\right],
\end{align}
where the Humbert function is defined in terms of Pochhammer symbols
$(x)_y = \Gamma(x{+}y)/\Gamma(x)$ as
\cite{bateman_vol1} % page 225, Eq. (24),
\begin{align}
  \label{ncr}
  \Psi_{(2)}(a,b,c; x,y)
  &= \sum_{m,n=0}^\infty\frac{(a)_{m+n}}{(b)_m (c)_{n}} x^m y^n,
\end{align}
which for equal arguments reduces to
$\Psi_{(2)}(a,a,a; x,y) = e^{x+y}\,{}_0F_1(a;xy)$, and so
\begin{align}
  \label{ncs}
  p(\bmy\cond \delta,\Delta,\gamma,K) %
  &= \frac{C}{
    (1{+}\delta^2)^{L/2}\,
    (1{+}\Delta^2)^{K/2}
  }\,
  \exp\left[
    -\frac{\chi^2}{2(1{+}\delta^2)}
    -\frac{(\gamma^2 {+} F_\smK^2)}{2(1{+}\Delta^2)}
  \right]
    {}_0F_1\left[\frac{K}{2}; \frac{\gamma^2 F_\smK^2}{4(1{+}\Delta^2)^2}\right].
\end{align}
Unlike the \AIC\ derivation, we have no reason to maintain the
distinction between noise and model prior variances and can set
$\delta = \Delta$, so the evidence reduces via Eq.~(\ref{pth}) to
\begin{align}
  \label{nct}
  p(\bmy\cond \Delta,\gamma,K) %
  &= C (1{{+}}\Delta^2)^{-N/2}
  \exp\left[-\frac{\gamma^2{+}z^2}{2(1{+}\Delta^2)}\right]
  {}_0F_1\left[\frac{K}{2}; \frac{\gamma^2 F_\smK^2}{4(1{+}\Delta^2)^2}\right].
\end{align}
If signal-to-noise ratios are known beforehand, $\gamma$ can be set to
a fixed number; otherwise, it must remain indeterminate and integrated
out. Aiming to have a maximally uniform but proper prior for $\gamma$,
we use a half-Gaussian with arbitrarily large variance $\sigma_\gamma^2$,
\begin{align}
  \label{ncu}
  p(\gamma\cond \sigma_\gamma) 
  &= \sqrt{\frac{2}{\sigma_\gamma^2\pi}}\exp\left[-\frac{\gamma^2}{2\sigma_\gamma^2}\right]
  \qquad 0 < \gamma < \infty,
\end{align}
yielding the evidence
\begin{align}
  \label{ncv}
  p(\bmy\cond \Delta,\sigma_\gamma,\hypo_\smK) 
  &= \int \drm\gamma\,p(\bmy\cond \Delta,\gamma,\hypo_\smK) \,p(\gamma\cond\sigma_\gamma)
  \\
  &= C \frac{(1{{+}}\Delta^2)^{-(N-1)/2}}{\sqrt{1{+}\sigma_\gamma^2{+}\Delta^2}}
  \exp\left[-\frac{z^2}{2(1{+}\Delta^2)}\right]
  {}_1F_1\left[\frac{1}{2}; \frac{K}{2}; 
    \frac{\sigma_\gamma^2 F_\smK^2}{2(1{+}\Delta^2)(1{+}\Delta^2{+}\sigma_\gamma^2)}\right]
  \nonumber
\end{align}
 and Bayes Factor
\begin{align}
  % \label{ncw}
  BF[K; N] 
  &= {}_1F_1\left[\frac{1}{2}; \frac{K}{2}; 
    \frac{\sigma_\gamma^2 F_\smK^2}{2(1{+}\Delta^2)(1{+}\Delta^2{+}\sigma_\gamma^2)}\right] 
  \biggl/ {}_1F_1\left[\frac{1}{2}; \frac{N}{2}; 
    \frac{\sigma_\gamma^2 z^2}{2(1{+}\Delta^2)(1{+}\Delta^2{+}\sigma_\gamma^2)}\right].
  \nonumber
\end{align}
We can now take the limit $\sigma_\gamma\rightarrow\infty$ to obtain
\begin{align}
  \label{ncx}
  BF[K; N] 
  &= {}_1F_1\left[\frac{1}{2}; \frac{K}{2}; \frac{ F_\smK^2}{2(1{+}\Delta^2)}\right] 
  \biggl/ 
  {}_1F_1\left[\frac{1}{2}; \frac{N}{2}; \frac{z^2}{2(1{+}\Delta^2)}\right].
\end{align}
The role of $\gamma$ has been to differentiate model and noise
parameter behaviour. For finite $\Delta$, integration over both
$\gamma$ in Eq.~(\ref{ncv}) and over $r$ in Eqs.~(\ref{ncc}) and
(\ref{ncp}) results, however, in redundancy which can safely be
eliminated by letting $\Delta\rightarrow 0$: unlike the \AIC, our
scales are set not by $\Delta$ but by $\gamma$ so we have no further
need for it. The Bayes Factor hence simplifies to
\begin{align}	
  \label{ncy}
  BF[K; N] 
  &= {}_1F_1\left[\frac{1}{2}; \frac{K}{2}; \frac{ F_\smK^2}{2}\right] 
  \biggl/ {}_1F_1\left[\frac{1}{2}; \frac{N}{2}; \frac{z^2}{2}\right].
\end{align}
Using the asymptotic properties of the
confluent hypergeometric distribution, \cite{bateman_vol1}, 
% p. 278, Eq. (3),
\begin{align}
  \label{ncz}
  {}_1F_1\left[a; c; x\right] \sim \frac{\Gamma[c]}{\Gamma[a]} e^{x}x^{a-c},
\end{align}
we obtain for large $N$ a robust version of the \BIC,
\begin{align}
  \label{ndc}
  \mathrm{\BIC_{(robust)}} = -2\log BF[K;N] 
  &\approx \chi^2 + (K-1)\log[F_\smK^2/2] - 2 \log \Gamma[K/2],
\end{align}
which for large $K$ reduces further to
\begin{align}
  \label{ndf}
  \mathrm{\BIC_{(robust)}} \simeq \chi^2 + K\log [F_\smK^2/K] + K.
\end{align}
The three new forms of the \BIC\ in Eqs.~(\ref{ncy}), (\ref{ncz}) and
(\ref{ndf}) are our central result.

We now show that, in the appropriate limits, the robust version approaches the
\AIC\ and \BIC.
The prior expectation value of $q^2$ for the Gamma Distribution
(\ref{ncj}) is $E[q^2] = L\delta^2/2$, while $E[r^2] = \gamma^2 + K
\Delta^2/2 $ for the Noncentral Gamma Distribution (\ref{nck}), so
that via Eq.~(\ref{ncl}) each parameter scales on average as
\begin{align}	
  \label{nde} 
  E[\hat{\beta}_j^2] 
  &\simeq \begin{cases} \Delta^2/2 + \gamma^2/K,  & j \leq K, \\
    \delta^2/2, & j > K.\end{cases}
\end{align}
As a result, the expectation value of the partial sum $F_j^2 =
\sum_{i=1}^j \hat{\beta}_i^2$ for $j=1,\ldots,N$ scales as
\begin{align}	
  E[F_j^2] 
  &\simeq  \sum_{i=1}^j E[\hat{\beta}_i^2] 
    = \begin{cases} 
      j \Delta^2/2 + j\gamma^2/K, & j < K, \\
    K\Delta^2/2 + \gamma^2 + (K-j)\delta^2/2,  & j \geq K.
  \end{cases}
\end{align}
In the \BIC\ limit, $\Delta^2 = N$ and $\delta=0$, so that $E[F_\smK^2/K] =
N/2 + \mbox{constant}$ and so for $N\gg K$, the robust \BIC\ becomes the \BIC\ up
to a constant. In the \AIC\ limit, $\Delta^2=\delta^2=1$ so that
$E[F_\smK^2/K] = 1 + \mathcal{O}(\gamma^2/K)$ and NIC reduces to approximately
$\chi^2 + K + \mathcal{O}(\gamma^2/K)$, close to the \AIC's $\chi^2+2K$.

\section{Results}
\label{sec:smm}
 
\noindent
To test the performance of the our robust \BIC, we present in this section a
numerical simulation, followed by semi-analytical estimates of the
salient quantities.

In the first part, we tested the success rate of information criteria in 
correctly identifying the number of parameters $S$ for competing models 
with varying parameter number $K$. Data sampling points $x_n$ were spread 
evenly over the interval $[0,\pi]$,
\begin{align}
  \label{smc}
  x_n &= \frac{(2n{-}1)\pi}{2N} \quad n = 1,\ldots,N,
\end{align}
and we generated $N=32$ data points throughout, setting the
experimental uncertainties to $\sigma_n=1$.
For each model $\hypo_\smsim$ constructed from $\Ksim$ simulation
parameters, data $\bmz(\Ksim) = (z_1(\Ksim),\ldots,z_\smN(\Ksim))$ was
generated as the sum of an ``ideal data'' term, a cosine series
\begin{align}
  \label{smd}
  f_k(x_n)
  &= \begin{cases} 
    \sqrt{\frac{1}{N}}& k=1, \\[8pt]
    \sqrt{\frac{2}{N}} \cos[(k{-}1)x_n] & k = 2,\ldots,\Ksim,
  \end{cases} 
\end{align}
whose amplitude $a$ was varied randomly by an additive term $b\phi_k$
with $\phi_k$ drawn from the standardised Gaussian distribution
$\phi_k\sim \mathcal{N}(0,1)$ and $b \geq 0$ an adjustable parameter.
Conceptually, $a$ represents the signal strength while $b$ controls
the variance of the signal. To simulate randomness associated with the
experimental uncertainty normally captured in $\sigma_n$, a second
random term $\varepsilon_n \sim \mathcal{N}(0,1)$ was added, so that
the 32 data points generated from the true parameter model
$\hypo_\smsim$ are
\begin{align}
  \label{sme}
  z_n(\Ksim) 
  &= \left[\sum_{k=1}^\Ksim f_k(x_n)\,(a + b \phi_k)\right] + \varepsilon_n(\Ksim).
\end{align}
For quenched values of $\phi_k$ and $\varepsilon_n$, one dataset was
generated for each $\Ksim = 1,2,\ldots N = 32$. All datasets were
efficiently computed in terms of $(N{\times}N)$-dimensioned matrices
\begin{align}
  \label{smf}
  \matD &= \matX_\smN(a\,\matI + b\,\matF)\matA +\matE,
\end{align}
where $\matD$ contains the $N$ column vectors $\bmz(\Ksim)$, one for
each $\Ksim$, matrix $\matX_\smN$ has elements
$(\matX_\smN)_{nk}=f_k(x_n)$, $\matI$ is the diagonal matrix, noise
matrix $\matF$ is diagonal with elements $\phi_k$ and $\matE$ contains
the $N^2$ gaussian random numbers $\varepsilon_n(\Ksim)$ with unit
variance. To limit the $k$-sum in Eq.~(\ref{sme}) to $\Ksim$, one must
include an upper-triangular matrix with components
$\matA_{\smK,\smsim} = \Theta(K,\Ksim) = 1$ for integers $K \geq \Ksim$
and 0 otherwise. 
To calculate $F_\smK^2$ and $\chi^2$ for given $\Ksim$ for use in the Bayes
Factor (\ref{ncy}) and elsewhere, we must modify the notation to keep
track of the ``true'' number of parameters $\Ksim$ to be compared to
the number of model parameters $K$. We therefore write
$\betamode_{\smK|\smsim}$ for the parameter mode of the model with $K$
parameters for data simulated from $\Ksim$ parameters; correspondingly
Eq.~(\ref{rsq}) becomes
$F_{\smK|\smsim}^2 = \betamode_{\smK|\smsim}^\tr\betamode_{\smK|\smsim}^{\,} =
\sum_{k=1}^K \hat{\beta}_{k|\smsim}^2$.

For model construction, we used the same cosine functions (\ref{smd})
used for data generation, and since the cosines form an orthonormal
system, the Hessian is diagonal,
$\matH_\smK = \matX_{\smK|\smsim}^\tr\matX_{\smK|\smsim} = \matI_\smK$, as
are the rotation and eigenvalue matrices, so that with
$\amode_{\smK|\smsim} = \matH_\smK^{-1}\matX_\smK^\tr\bmz(\Ksim)$ the
mode simplifies to
\begin{align}
  \label{smi}
  \betamode_{\smK|\smsim} 
  &= \matL_\smK^{\!\!1/2}\matS_\smK^\tr\amode_\smK
  = \matX_\smK^\tr \bmz(\Ksim).
\end{align}
The $N{\times}K$ design matrix $\matX_\smK$ is augmented by means of a
projector $\matP_\smK$ which contains 1's along its first $K$ diagonal
elements and 0 elsewhere; the truncated design matrix
$\matX_\smK = \matX_\smN\matP_\smK$ %
then contains zeros in the last $N{-}K$ columns of the $N{\times}N$
matrix and $f_k(x_n)$ elsewhere. %
The $N{\times}N$ matrix of modes $\matB$ with elements
$\matB_{\smK,\smsim} = \hat{\beta}_{\smK|\smsim}$ is then compactly
represented as
\begin{align}
  \label{smj}
  \matB &= \matX_\smN^\tr \matP_\smK \matD.
\end{align}
Note that this matrix formulation is possible only for orthogonal
basis functions; for nonorthogonal cases, the Hessian and its
eigensystem must be recalculated for every $K$.

We now turn to the test results. Given a dataset with $\Ksim$
parameters, the $K$ value which minimises the AIC, BIC or the robust version 
in its Eq.~(\ref{ndf}) form is deemed a success if that $K$ correctly
matches the data's $\Ksim$. In Fig.~\ref{fig:success}, we plot the
number of successes as percentages of $2^{16}$ repetitions of 32
datasets as described above as a function of $\Ksim$ for four
information criteria. The upper panel displays results for weak-signal
data generated with $a=1,b=1$, while the strong-signal data shown in
the lower panel used $a=5,b=1$.  Red diamonds represent the robust $\BIC$
success rate, green circles the corresponding \AIC\ success rate and
black triangles the \BIC. Also shown as blue squares is the success
rate of the corrected \AIC,  \cite{george2000calibration}.
 
As expected, the \BIC performs poorly in the weak-signal environment 
where the models are badly specified but very well for strong signals, 
where the models are more successful.  While the \AIC\ is more successful in
the weak-signal case but underperforms the \BIC\ for strong signals.  It
is not a surprise that the corrected \AIC, which was designed for a particular
subset of data scenarios, does very well in the mid-range of the
strong-signal case but fails badly otherwise.\footnote{The corrected \AIC
  corrects the \AIC formula for small $N$ and therefore small
  $\Ksim$. In effect, this improves the \AIC for strong-signal cases,
  but destroys its performance for the weak-signal case and larger
  $\Ksim$.}  By contrast, the robust \BIC matches or exceeds the performance
of all other information criteria in both the strong and weak signal
scenarios.  The exact \BIC\ result (\ref{ncy}) and the Eqs.~(\ref{ndf})
differ by less than one percent.

The general increase in success rates for $\Ksim$ near the simulation
lower limit 1 and upper limit 32 are easily understood because there
are fewer alternatives to $\Ksim$ at these edges. While this rise will
persist for small $\Ksim$, it will for large $\Ksim$ shift with
increasing $N$ and is therefore a nonpersistent ``boundary effect''.
%%%%%%%%%%%%%%%%%%%%%%%%%%%%%%%%%%%%%%%%%%%%%%%%%%%%%%%%%%%%%%%%%%%%%%%%%%%%%%
\begin{figure}
  \centering
  \begin{tikzpicture}[every node/.style={draw, inner sep=0pt,minimum size=1.5mm}]
    \newcommand{\Xscale}{0.40}
    \newcommand{\Yscale}{8}    	
    % Axes + Title
    \node[draw=none] at (15*\Xscale,1.05*\Yscale) {\textbf{Weak Signal}};
    \draw[ultra thick] (0,-0.2) -- (32*\Xscale,-0.2);
    \draw[ultra thick] (0,-0.2) -- (0,1*\Yscale);
    % Tikz
    \foreach \x in {1,3,...,31}
    \draw (\x*\Xscale,-0.2) -- (\x*\Xscale,-0.35)
    node[draw=none,anchor=north] {\small \textbf{\x}};
    \foreach \x in {2,...,32}
    \draw (\x*\Xscale,-0.2) -- (\x*\Xscale,-0.35)
    node[draw=none,anchor=north] {};
    \foreach \y in {0,10,...,100}
    \draw (0,\y/100*\Yscale) -- (-0.15,\y/100*\Yscale) 
    node[draw=none,anchor=east] {\small \textbf{\y\%}}; 
    % labels      
    \node[draw=none,below=0.8cm] at (16*\Xscale,0) {\textbf{Simulation parameters $\Ksim$}};
    \node[draw=none,left=1.6cm,rotate=90] at (0,0.7*\Yscale) {\textbf{Percentage of successes}};
    % AIC    	
    \node[circle,line width=0mm,fill=green] (AIC-legend1) at (27*\Xscale,0.9*\Yscale) {};
    \node[circle,line width=0mm,fill=green] (AIC-legend2) at (28*\Xscale,0.9*\Yscale) {};
    \draw[green] (AIC-legend1) -- (AIC-legend2);
    \node[draw=none] at (30*\Xscale,0.9*\Yscale) {AIC};
    % AIC-data
    \foreach \x/\y in {1/0.306,2/0.257,3/0.229,4/0.217,5/0.207,6/0.2,7/0.193,8/0.192,9/0.187,    
      10/0.182,11/0.185,12/0.182,13/0.178,14/0.177,15/0.176,16/0.177,17/0.174,18/0.172,
      19/0.175,20/0.176,21/0.174,22/0.171,23/0.171,24/0.175,25/0.173,26/0.173,27/0.174,
      28/0.175,29/0.177,30/0.189,31/0.201,32/0.236}
    {
      \node[circle,line width=0mm,fill=green] (Aic\x) at (\x*\Xscale,\y*\Yscale) {};
    }
    \foreach \x [evaluate=\x as \y using int(\x+1)] in {1,...,31} 
    {
      \draw[green] (Aic\x) -- (Aic\y);
    }
    % BIC
    \node[regular polygon,regular polygon sides=3,line width=0mm,inner sep=0.3mm,fill=black] (BIC-legend1) at (27*\Xscale,0.85*\Yscale) {};
    \node[regular polygon,regular polygon sides=3,line width=0mm,inner sep=0.3mm,fill=black] (BIC-legend2) at (28*\Xscale,0.85*\Yscale) {};
    \draw[black] (BIC-legend1) -- (BIC-legend2);
    \node[draw=none] at (30*\Xscale,0.85*\Yscale) {BIC};
    % % % % % 
    \foreach \x/\y in 	
    {1/0.27,2/0.19,3/0.149,4/0.125,5/0.108,6/0.096,7/0.086,8/0.078,9/0.072,    
      10/0.066,11/0.063,12/0.059,13/0.055,14/0.052,15/0.048,16/0.048,17/0.044,18/0.041,
      19/0.039,20/0.039,21/0.037,22/0.035,23/0.033,24/0.032,25/0.031,26/0.029,27/0.029,
      28/0.028,29/0.026,30/0.025,31/0.025,32/0.026}{
      \node[regular polygon,regular polygon sides=3,line width=0mm,inner sep=0.33mm,fill=black] (Bic\x) at (\x*\Xscale,\y*\Yscale) {};
    }
    \foreach \x [evaluate=\x as \y using int(\x+1)] in {1,...,31} 
    {
      \draw[black] (Bic\x) -- (Bic\y);
    }
    % NIC
    \node[diamond,line width=0mm,fill=red] (NIC-legend1) at (27*\Xscale,0.80*\Yscale) {};
    \node[diamond,line width=0mm,fill=red] (NIC-legend2) at (28*\Xscale,0.80*\Yscale) {};
    \draw[red] (NIC-legend1) -- (NIC-legend2);
    \node[draw=none] at (30*\Xscale,0.80*\Yscale) {NIC};    	      	
    % % % 
    \foreach \x/\y in 	
    {1/0.259,2/0.22,3/0.201,4/0.195,5/0.193,6/0.19,7/0.187,8/0.187,9/0.187,    
      10/0.184,11/0.187,12/0.187,13/0.186,14/0.185,15/0.185,16/0.189,17/0.188,18/0.187,
      19/0.193,20/0.191,21/0.192,22/0.192,23/0.193,24/0.198,25/0.197,26/0.2,27/0.204,
      28/0.209,29/0.216,30/0.232,31/0.258,32/0.318}
    {
      \node[diamond,line width=0mm,fill=red] (Mic\x) at (\x*\Xscale,\y*\Yscale) {};
    }
    \foreach \x [evaluate=\x as \y using int(\x+1)] in {1,...,31} 
    {
      \draw[red] (Mic\x) -- (Mic\y);
    }
    % AICc
    \node[rectangle,line width=0mm,fill=blue] (AICc-legend1) at (27*\Xscale,0.75*\Yscale) {};
    \node[rectangle,line width=0mm,fill=blue] (AICc-legend2) at (28*\Xscale,0.75*\Yscale) {};
    \draw[blue] (AICc-legend1) -- (AICc-legend2);
    \node[draw=none] at (30*\Xscale,0.75*\Yscale) {AICc};	    
    % AICc-data
    \foreach \x/\y in 	
    {1/0.328,2/0.287,3/0.285,4/0.21,5/0.176,6/0.155,7/0.145,8/0.106,9/0.086,    
      10/0.072,11/0.051,12/0.037,13/0.022,14/0.015,15/0.007,16/0.003,17/0.002,18/0,
      19/0,20/0,21/0,22/0,23/0,24/0,25/0,26/0,27/0,
      28/0,29/0,30/0,31/0,32/0}
    {
      \node[rectangle,line width=0mm,fill=blue] (Aicc\x) at (\x*\Xscale,\y*\Yscale) {};
    }
    \foreach \x [evaluate=\x as \y using int(\x+1)] in {1,...,31} 
    {
      \draw[blue] (Aicc\x) -- (Aicc\y);
    }
  \end{tikzpicture}
  %%%%%%%%%%%%%%%%%%%%%%%%%%%%%%%%%%%%%%%%%%%%%%%%%%%%%%%%%%%%%%%%%%%%%%%%%%%%%% 
  \par\vspace*{10mm}
  %%%%%%%%%%%%%%%%%%%%%%%%%%%%%%%%%%%%%%%%%%%%%%%%%%%%%%%%%%%%%%%%%%%%%%%%%%%%%% 
  \begin{tikzpicture}[every node/.style={draw, inner sep=0pt,minimum size=1.5mm}]
    \newcommand{\Xscale}{0.40}
    \newcommand{\Yscale}{8}    	
    % Axes + Title
    \node[draw=none] at (15*\Xscale,1.05*\Yscale) {\textbf{Strong Signal}};
    \draw[ultra thick] (0,-0.2) -- (32*\Xscale,-0.2);
    \draw[ultra thick] (0,-0.2) -- (0,1*\Yscale);
    % Tikz
    \foreach \x in {1,3,...,31}
    \draw (\x*\Xscale,-0.2) -- (\x*\Xscale,-0.35)
    node[draw=none,anchor=north] {\small \textbf{\x}};
    \foreach \x in {2,...,32}
    \draw (\x*\Xscale,-0.2) -- (\x*\Xscale,-0.35)
    node[draw=none,anchor=north] {};
    \foreach \y in {0,10,...,100}
    \draw (0,\y/100*\Yscale) -- (-0.15,\y/100*\Yscale) 
    node[draw=none,anchor=east] {\small \textbf{\y\%}}; 
    % labels      
    \node[draw=none,below=0.8cm] at (16*\Xscale,0) {\textbf{Simulation parameters $\Ksim$}};
    \node[draw=none,left=1.6cm,rotate=90] at (0,0.7*\Yscale) {\large \textbf{Percentage of successes}};
    % AIC    	
    \node[circle,line width=0mm,fill=green] (AIC-legend1) at (2*\Xscale,0.2*\Yscale) {};
    \node[circle,line width=0mm,fill=green] (AIC-legend2) at (3*\Xscale,0.2*\Yscale) {};
    \draw[green] (AIC-legend1) -- (AIC-legend2);
    \node[draw=none] at (5*\Xscale,0.2*\Yscale) {AIC};
    %
    % AIC-data
    \foreach \x/\y in {1/0.707,2/0.708,3/0.707,4/0.71,5/0.705,6/0.708,7/0.708,8/0.709,9/0.709,    
      10/0.709,11/0.708,12/0.71,13/0.71,14/0.708,15/0.709,16/0.707,17/0.713,18/0.707,
      19/0.709,20/0.712,21/0.71,22/0.715,23/0.712,24/0.719,25/0.719,26/0.727,27/0.734,
      28/0.736,29/0.753,30/0.783,31/0.841,32/0.994}
    {
      \node[circle,line width=0mm,fill=green] (Aic\x) at (\x*\Xscale,\y*\Yscale) {};
    }
    \foreach \x [evaluate=\x as \y using int(\x+1)] in {1,...,31} 
    {
      \draw[green] (Aic\x) -- (Aic\y);
    }
    % BIC
    \node[regular polygon,regular polygon sides=3,inner sep=0.33mm,line width=0mm,fill=black] (BIC-legend1) at (2*\Xscale,0.15*\Yscale) {};
    \node[regular polygon,regular polygon sides=3,inner sep=0.33mm,line width=0mm,fill=black] (BIC-legend2) at (3*\Xscale,0.15*\Yscale) {};
    \draw[black] (BIC-legend1) -- (BIC-legend2);
    \node[draw=none] at (5*\Xscale,0.15*\Yscale) {BIC};
    % % % % % BIC
    \foreach \x/\y in 	
    {1/0.906,2/0.904,3/0.904,4/0.906,5/0.906,6/0.902,7/0.903,8/0.905,9/0.904,    
      10/0.907,11/0.904,12/0.904,13/0.906,14/0.905,15/0.905,16/0.905,17/0.906,18/0.904,
      19/0.905,20/0.905,21/0.904,22/0.904,23/0.903,24/0.906,25/0.906,26/0.906,27/0.905,
      28/0.905,29/0.906,30/0.911,31/0.927,32/0.987}
    {
      \node[regular polygon,regular polygon sides=3,inner sep=0.33mm,line width=0mm,fill=black] (Bic\x) at (\x*\Xscale,\y*\Yscale) {};
    }
    \foreach \x [evaluate=\x as \y using int(\x+1)] in {1,...,31} 
    {
      \draw[black] (Bic\x) -- (Bic\y);
    }
    % NIC
    \node[diamond,line width=0mm,fill=red] (NIC-legend1) at (2*\Xscale,0.10*\Yscale) {};
    \node[diamond,line width=0mm,fill=red] (NIC-legend2) at (3*\Xscale,0.10*\Yscale) {};
    \draw[red] (NIC-legend1) -- (NIC-legend2);
    \node[draw=none] at (5*\Xscale,0.10*\Yscale) {NIC};    	      	
    % % % % % % 
    \foreach \x/\y in 	
    {1/0.884,2/0.892,3/0.897,4/0.901,5/0.902,6/0.898,7/0.9,8/0.902,9/0.901,    
      10/0.903,11/0.901,12/0.901,13/0.904,14/0.902,15/0.902,16/0.901,17/0.903,18/0.902,
      19/0.901,20/0.902,21/0.902,22/0.902,23/0.9,24/0.903,25/0.903,26/0.904,27/0.903,
      28/0.903,29/0.903,30/0.909,31/0.925,32/0.987}
    {
      \node[diamond,line width=0mm,fill=red] (Mic\x) at (\x*\Xscale,\y*\Yscale) {};
    }
    \foreach \x [evaluate=\x as \y using int(\x+1)] in {1,...,31} 
    {
      \draw[red] (Mic\x) -- (Mic\y);
    }
    % AICc
    \node[rectangle,line width=0mm,fill=blue] (AICc-legend1) at (2*\Xscale,0.05*\Yscale) {};
    \node[rectangle,line width=0mm,fill=blue] (AICc-legend2) at (3*\Xscale,0.05*\Yscale) {};
    \draw[blue] (AICc-legend1) -- (AICc-legend2);
    \node[draw=none] at (5*\Xscale,0.05*\Yscale) {AICc};	    
    % AICc-data
    \foreach \x/\y in 	
    {1/0.759,2/0.792,3/0.878,4/0.878,5/0.881,6/0.888,7/0.932,8/0.929,9/0.933,    
      10/0.955,11/0.948,12/0.958,13/0.949,14/0.958,15/0.932,16/0.932,17/0.917,18/0.877,
      19/0.824,20/0.765,21/0.666,22/0.521,23/0.329,24/0.153,25/0.031,26/0.001,27/0,
      28/0,29/0,30/0,31/0,32/0}
    {
      \node[rectangle,line width=0mm,fill=blue] (Aicc\x) at (\x*\Xscale,\y*\Yscale) {};
    }
    \foreach \x [evaluate=\x as \y using int(\x+1)] in {1,...,31} 
    {
      \draw[blue] (Aicc\x) -- (Aicc\y);
    }
  \end{tikzpicture}
  \caption{Comparison of success rates over $2^{16}$ datasets of
    information criteria for ``weak signal'' and ``strong signal''
    scenarios. Given data generated with $\Ksim$ parameters, a
    particular information criterion is deemed successful if it picks
    the model with the correct number of parameters, $K=\Ksim$.}
  \label{fig:success}
\end{figure}
%%%%%%%%%%%%%%%%%%%%%%%%%%%%%%%%%%%%%%%%%%%%%%%%%%%%%%%%%%%%%%%%%%%%%%%%%%%%%%

In the second test, we utilise the simple linear system of
Eqs.~(\ref{smf})--(\ref{smj}) to obtain analytical estimates of the
squared signal and noise for a detailed but statistically approximate
analysis of the shapes and sizes of the criteria's $K$ vs $\Ksim$
curves.  Because $\matF^\tr\matF \simeq \matI$ and
$\var(\varepsilon)=1$, the squared data vectors scale approximately as
\begin{align}
  \label{smg}
  z^2(\Ksim) 
  = \sum_n z_n^2(\Ksim)
  &= \text{diag}(\matD^T\matD)
  \ \simeq\ N + (a^2+b^2)\Ksim.
\end{align} 
The squared signal is obtained from the diagonal elements of the
squared mode matrix, $F_{\smK|\smsim}^2 =
[\matB^\tr\matB]_{\smsim,\smsim}$. Inserting the explicit simulation
model (\ref{smf}) results in an approximate estimate of
\begin{align}
  \label{smk}
  F_{\smK|\smsim}^2 
  &\simeq (a^2 + b^2)\matA^{\!\tr} \matP_\smK \matA + \matE^\tr \matP_\smK \matE
   = (a^2 + b^2)\min(K,\Ksim) + K
\end{align}
while the squared data vector and squared noise vector are obtained from
\begin{align}
  \label{sml}
  z^2(S)
  &= (\matD^\tr\matD)_{\smsim,\smsim} = (a^2+b^2)\Ksim + N \\
  \label{smm}
  \chi^2_{\smK|\smsim} 
  &= z^2(S) - F_{\smK|\smsim}^2 \nonumber\\
  &\simeq N - K + (a^2 + b^2)[\Ksim - \min(K,\Ksim)]\nonumber\\
  &= \left\{
    \begin{array}{ll}
      N-K + (a^2+b^2)(\Ksim-K) & \text{ for } K < \Ksim,\\
      N-K  & \text{ for } K \geq \Ksim.\\
    \end{array}
    \right.
\end{align}
%%%%%%%%%%%%%%%%%%%%%%%%%%%%%%%%%%%%%%%%%%%%%%%%%%%%%%%%%%%%%%%%%%%%%%%%%%%%%%%
\begin{figure}[h]
  \centering
  \begin{tikzpicture}[every node/.style={draw, inner sep=0pt,minimum size=2.0mm}]
    \newcommand{\Xscale}{0.65}
    \newcommand{\Yscale}{0.12}
    % \newcommand{\Xscale}{0.7}
    % \newcommand{\Yscale}{0.15}
    % % % Legends
    \node[circle,line width=0mm,fill=green] (AIC-legend1) at (6.5*\Xscale,75*\Yscale) {};
    \node[circle,line width=0mm,fill=green] (AIC-legend2) at (7.5*\Xscale,75*\Yscale) {};
    \draw[green] (AIC-legend1) -- (AIC-legend2);
    \node[draw=none] at (8.5*\Xscale,75*\Yscale) {\textbf{AIC}};
    % % % % 
    \node[regular polygon,regular polygon sides=3,inner sep=0.45mm,line width=0mm,fill=black] (BIC-legend1) at (10.0*\Xscale,75*\Yscale) {};
    \node[regular polygon,regular polygon sides=3,inner sep=0.45mm,line width=0mm,fill=black] (BIC-legend2) at (11.0*\Xscale,75*\Yscale) {};
    \draw[black] (BIC-legend1) -- (BIC-legend2);
    \node[draw=none] at (12.0*\Xscale,75*\Yscale) {\textbf{BIC}};
    % % % % 
    \node[diamond,line width=0mm,fill=red] (NIC-legend1) at (6.5*\Xscale,70*\Yscale) {};
    \node[diamond,line width=0mm,fill=red] (NIC-legend2) at (7.5*\Xscale,70*\Yscale) {};
    \draw[red] (NIC-legend1) -- (NIC-legend2);
    \node[draw=none] at (8.5*\Xscale,70*\Yscale) {\textbf{NIC}};
    % % % % 
    \node[rectangle,line width=0mm,fill=blue] (Chi-legend1) at (10.0*\Xscale,70*\Yscale) {};
    \node[rectangle,line width=0mm,fill=blue] (Chi-legend2) at (11.0*\Xscale,70*\Yscale) {};
    \draw[blue] (Chi-legend1) -- (Chi-legend2);
    \node[draw=none] at (12.0*\Xscale,70*\Yscale) {$\bmchi^2$};
    % Axes + Title
    \draw[ultra thick] (3.5*\Xscale,15*\Yscale) -- (12.5*\Xscale,15*\Yscale);
    \draw[ultra thick] (3.5*\Xscale,15*\Yscale) -- (3.5*\Xscale,80*\Yscale);
    % % % Labels
    \node[draw=none,below=0.8cm] at (8*\Xscale,15*\Yscale) {\textbf{Model parameters $K$}};
    %\node[draw=none,left=1.6cm,rotate=90] at (3.5*\Xscale,50*\Yscale) {$\bmchi$};   
    % % % % Tikz
    \foreach \x in {4,...,12}
    \draw (\x*\Xscale,15*\Yscale) -- (\x*\Xscale,15*\Yscale-0.2)
    node[draw=none,anchor=north] {\small \textbf{\x}};
    % \foreach \y in {15,22.5,...,75}
    \foreach \y in {15,20,...,75}
    \draw (3.5*\Xscale,\y*\Yscale) -- (3.5*\Xscale-0.2,\y*\Yscale)
    node[draw=none,anchor=east] {\small \textbf{\y}};
    % Data
    \foreach \x/\y/\z in {4/32.0/0.693,5/30.0/0.693,6/28.0/0.693,7/26.0/0.693,8/24.0/0.693,9/23.0/0.636,10/22.0/0.588,11/21.0/0.547,12/20.0/0.511}
    {
      \node[circle,line width=0mm,fill=green] (AIC\x) at (\x*\Xscale,\y*\Yscale+2*\x*\Yscale) {};
      \node[regular polygon,regular polygon sides=3,inner sep=0.45mm,line width=0mm,fill=black] (BIC\x) at (\x*\Xscale,\y*\Yscale+3.47*\x*\Yscale) {};
      \node[diamond,line width=0mm,fill=red] (NIC\x) at (\x*\Xscale,\y*\Yscale + \x*\Yscale + \z*\Yscale*\x) {};
      \node[rectangle,line width=0mm,fill=blue] (Chi\x) at (\x*\Xscale,\y*\Yscale) {};
    }
    \foreach \x[evaluate=\x as \y using int(\x+1)] in {4,...,11} {
      \draw[green] (AIC\x) -- (AIC\y);
      \draw[black] (BIC\x) -- (BIC\y);
      \draw[red] (NIC\x) -- (NIC\y);
      \draw[blue] (Chi\x) -- (Chi\y);
    }
    \foreach \x/\y/\z in
    {4/64.0/2.3,5/54.0/2.3,6/44.0/2.3,7/34.0/2.3,8/24.0/2.3,9/23.0/2.2,10/22.0/2.1,11/21.0/2.0,12/20.0/1.95}
    {
      \node[circle,line width=0mm,fill=green] (AIC2\x) at (\x*\Xscale,\y*\Yscale+2*\x*\Yscale) {};
      \node[regular polygon,regular polygon sides=3,inner sep=0.45mm,line width=0mm,fill=black] (BIC2\x) at (\x*\Xscale,\y*\Yscale+3.47*\x*\Yscale) {};
      \node[diamond,line width=0mm,fill=red] (NIC2\x) at (\x*\Xscale,\y*\Yscale + \x*\Yscale + \z*\Yscale*\x) {};
      \node[rectangle,line width=0mm,fill=blue] (Chi2\x) at (\x*\Xscale,\y*\Yscale) {};
   }
   \foreach \x[evaluate=\x as \y using int(\x+1)] in {4,...,11} {
     \draw[green] (AIC2\x) -- (AIC2\y);
     \draw[black] (BIC2\x) -- (BIC2\y);
     \draw[red] (NIC2\x) -- (NIC2\y);
     \draw[blue] (Chi2\x) -- (Chi2\y);
   }
   \end{tikzpicture} 
   \caption{Evolution of information criteria for model parameter
     numbers $K$ versus true parameter number $\Ksim=8$, for strong
     signal using $a=3,b=0$ (upper curves) and weak signal using
     $a=1,b=0$ (lower curves). Also shown is the evolution of
     $\chi^2$. For $K>\Ksim$, the AIC and BIC do not distinguish between
     strong and weak signal cases.}
    \label{fig:sim}
\end{figure}
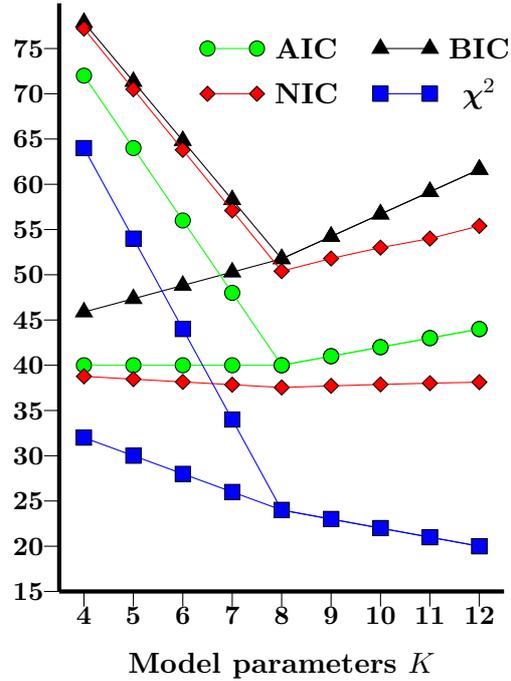
%%%%%%%%%%%%%%%%%%%%%%%%%%%%%%%%%%%%%%%%%%%%%%%%%%%%%%%%%%%%%%%%%%%%%%%%%%%%%%%
These expressions provide instructive, if approximate, insights into
the behaviour of the information criteria as a function of model
parameter number $K$. Fig.~\ref{fig:sim} illustrates by example the
shapes of the minima as functions of $K$ of the simplest robust \BIC form
(\ref{ndf}) as well as the \AIC and \BIC for fixed $\Ksim=8,b=0$ and
strong-signal $a=3$ and weak-signal $a=1$ scenarios.
The \AIC and \BIC do not depend on $F_\smK^2$ but only on $\chi^2$, which
exhibits the well-known behaviour of steadily decreasing with
$K$. Upper and lower branches of these curves denote the strong-signal
and weak-signal cases respectively.
Based on $\chi^2$ and the simple penalty terms, the \AIC and \BIC
both exhibit a reasonably strong minimum at $K{=}\Ksim$ for $a=3$; for
the weak-signal $a{=}1$, however, the \AIC remains flat while the \BIC
has no minimum at all.  This is reflected in the low \BIC success rate
in Fig.~\ref{fig:success}.
Since $\chi^2$ becomes independent of $a$ for $K \geq \Ksim$, the \AIC
and \BIC do not distinguish between strong and weak scenarios in that
region. 

The robust \BIC, by contrast, is sensitive to the squared signal $F_\smK^2$, which
lifts the degeneracy between strong and weak signal for $K\geq\Ksim$.
Like the \AIC and \BIC, the robust \BIC has no trouble identifying $K=S$ for
strong signal. For weak signal, it exhibits a minimum at the
correct answer, albeit a shallow one. Shallow minima reflect, of
course, the inherent uncertainty regarding the signal or noise
character of the data.
Details of Figure \ref{fig:sim} and its discussion are, of course,
specific to the model and numbers used and of illustrative value only.

\section{Discussion and conclusions}
\label{sec:dsc}

The robust version of the $\BIC$ introduced in this paper is based
on three simple but novel ideas. Firstly, we have expanded Akaike's
original argument for a larger model space into a model space plus a
fully-fledged noise space which together partition the entire data
space.  The resulting symmetries and scale behaviour of model and
noise space provide a surprisingly unified and indeed beautiful
framework for linear regression.

Secondly, building on the insight of earlier work
\cite{dekock2017bayesian}, we posit that both model parameter and
noise parameter spaces should be projected onto a radial coordinate on
the respective hypersphere. Unlike \cite{dekock2017bayesian}, however,
we now have not one but two hyperspheres reflecting the separate
symmetries and scales of the model and noise spaces.

The third insight is that the crucial difference between model and
noise parameters lies not in the scales $\delta$ and $\Delta$ ---
indeed we set these equal and eventually even set $\Delta = 0$ --- but
in explicitly taking into account that the maximum-likelihood
parameter vector's magnitude ${\parallel}\betamode_\smK{\parallel}$
must, by the very definition of ``signal'', be significantly nonzero,
while ${\parallel}\betamode_\smL{\parallel}\simeq 0$ for noise.  This
results in a Gamma Distribution for noise parameters arising from
projection of a zero-mode Gaussian on the one hand, and a noncentral
Gamma Distribution for model parameters arising from projection of a
nonzero-mode Gaussian.

Together, these three insights have allowed us to calculate Bayes
Factors for model comparison in closed form and construct a robust version 
of the \BIC\ which extends the robustness the \AIC\ has against model misspecification to the \BIC. Unlike the latter, the robust \BIC depends explicitly on the
squared signal strength $F_\smK^2$, and as $F_\smK^2$ approaches the weak or
the strong signal limit, the robust \BIC\ correspondingly approaches the \AIC
and \BIC cases as limiting forms.

The noncentrality parameter $\gamma$ as a measure of signal strength
appears to be the essence of the difference between signal and noise.
Where the signal-to-noise ratio is known beforehand, $\gamma$ can be
set to a fixed number or restricted to a limited interval. In the
general case of unknown signal-to-noise ratio, however, it is better
to integrate $\gamma$ over all possible values as implemented here.

We conclude with a few general remarks.
Naturally, the scope of the numerical results presented here is
limited, and this robust version of the \BIC\ should be tested and possibly improved when
applied to a diversity of data scenarios.  The present results should
also be extended from the fixed experimental uncertainties $\bmsigma$
to variable $\sigma$.
The analysis was done in the context of linear regression and should strictly speaking be used only in that context. The degree of success for nonlinear situations cannot be
estimated or guaranteed within the present framework.
Our derivations presume that there is only one model per $K$. This
limited approach is easily generalised to include more than one model
for a given $K$ using, for example, indicator vectors as set out in
\cite{Liang2008}.
\\  

\noindent\textbf{Acknowledgements}\\
This work was supported in part by the South African National Research
Foundation.

%%%%%%%%%%%%%%%%%%%%%%%%%%%%%%%%%%%%%%%%%%%%%%%%%%%%%%%%%%%%%%%%%%%%%%%%%%%%%
\bibliography{dekockeggers-submitv2} 

\end{document}